\documentclass[twocolumn, a4paper]{article}
\usepackage{graphicx}
\usepackage{subcaption}
\usepackage{authblk}
\usepackage{xcolor}
\usepackage{braket}
\usepackage{amsmath}
\usepackage{siunitx}
\usepackage[export]{adjustbox}
\usepackage{mathtools}

\usepackage{enumitem}
\setlist[enumerate,1]{label={(\roman*)}}

\usepackage[
    style=ieee,
    citestyle=ieee-comp,
    url=false,
]{biblatex}
\usepackage[hidelinks]{hyperref}
\usepackage{orcidlink}

\newcommand{\erado}{\href{https://github.com/oqc-community/erado}{\texttt{erado}}}

\newcommand{\twodots}{\mathinner{\ldotp \ldotp}}

\newcommand{\gate}[1]{\mathrm{#1}}

\newcommand{\quantumop}[1]{\mathcal{#1}}

\newcommand{\complexity}[1]{\mathsf{#1}}

\DeclareMathOperator{\vecspan}{span}

\DeclareMathOperator{\trace}{tr}

\DeclareMathOperator*{\argmax}{arg\,max}

\makeatletter
\DeclareRobustCommand\ket[1]{%
  \@ifnextchar\bra{\k@t{#1}\!}{\k@t{#1}}%
}
\newcommand\k@t[1]{{|{#1}\rangle}}
\makeatother

\DeclarePairedDelimiter\abs{\lvert}{\rvert}

\newcommand{\middlemid}{\;\middle|\;}

\DeclareSIUnit{\quop}{quop}

\title{The limits of erasure-based postselection for quantum error mitigation}
\author[]{Sam J.\ Griffiths\thanks{sgriffiths@oqc.tech} \orcidlink{0000-0001-7211-4339}}
\author[]{Jamie Friel\thanks{jfriel@oqc.tech}}
\author[]{Brian Vlastakis\thanks{bvlastakis@oqc.tech}}
\affil[]{Oxford Quantum Circuits (OQC), United Kingdom, RG2 9LH}

\date{June 2026}

\begin{document}

\maketitle

\begin{abstract}
\small
In both classical and quantum error correction, heralded erasures are known to be easier to tolerate than unheralded general stochastic errors. Whilst an established benefit of loss-dominant quantum architectures such as photonic qubits, this fact has received renewed interest, with a pivot towards reconstructing other architectures to be erasure-dominant, such as dual-rail transmons. This work investigates exploiting these `erasure qubits' in the near term by using postselection as a technique for error mitigation, wherein circuit shots detecting any erased qubits are discarded from the computational ensemble and repeated. Firstly, we outline a numerical framework for representing circuit-level erasure noise and present `erado', an open-source library capable of simulating erasure noise and postselection. Secondly, we investigate the effects of both erasure noise and noise in the erasure checks themselves on the quantum Fourier transform (QFT), in the additional presence of gate depolarising noise. A worked example is provided of postselection fully mitigating against the erasure channel for erasure check error rates less than 3.0\%. We also show how a postselected dual-rail system can surpass a fundamental noise floor at the kiloquop scale where a comparable single-rail system cannot, justifying this approach in the NISQ regime before (and, perhaps, combined with) the practical arrival of QEC.
\end{abstract}

\section{Introduction}

A central challenge in many quantum computing architectures -- and especially in superconducting qubits -- is the presence of energy relaxation processes, commonly characterised by the $T_1$ lifetime \cite{burnettDecoherenceBenchmarkingSuperconducting2019}. In conventional architectures, this decay mechanism manifests as an unheralded amplitude damping channel, introducing stochastic errors that directly contribute to logical failure rates \cite{nielsenQuantumComputationQuantum2010}; the error mechanism is \emph{unheralded} in that we cannot directly know if or where discrete stochastic errors are induced, requiring algorithmic inference via error-correcting codes \cite{macwilliamsTheoryErrorcorrectingCodes1977, calderbankGoodQuantumErrorcorrecting1996, dennisTopologicalQuantumMemory2002}. This places stringent requirements on coherence, calibration, and ultimately the overhead required for quantum error correction (QEC).

In this work, we consider an approach based on erasure-aware qubit encodings, in which dominant relaxation events are heralded (that is, can be detected) and conditionally removed from the computational ensemble, i.e.\ \emph{postselection} \cite{aaronsonQuantumComputingPostselection2005}. In a \emph{dual-rail} encoding, two physical modes are used to represent a single logical qubit, enabling the detection of prevailing loss mechanisms through a single end-of-line measurement \cite{kokLinearOpticalQuantum2007, willsErrordetectedCoherenceMetrology2025}. This provides a practical route to mitigating the dominant source of noise in such systems, not only in their fault-tolerant future when combined with QEC, but also in the noisy intermediate-scale (NISQ) regime, wherein active QEC remains challenging for current hardware capabilities.

In Section~\ref{sec:background}, technical background is introduced: we formally define the dual-rail codespace, describe an erasure noise model appropriate for relaxation-dominant systems, explain the propagation of erasures through circuits, and analyse the role of imperfect erasure detection in postselection.

In Section~\ref{sec:numerical-techniques}, we overview our simulational techniques developed for benchmarking the effects of noise parameters upon the performance and cost of erasure-based postselection. We have published this codebase as \erado{}, an open-source Qiskit-based \cite{javadi-abhariQuantumComputingQiskit2024} library providing tools for the simulation of erasure noise and postselection with arbitrary quantum circuits.\footnote{\url{https://github.com/oqc-community/erado}}

In Section~\ref{sec:results}, we demonstrate these techniques in a series of results using the quantum Fourier transform (QFT) as a case study; in particular, we explore how the overhead of postselection scales with the size of the circuit, as well as how the mean fidelity with and without postselection is affected by erasure noise, erasure check noise and gate depolarising noise. Finally, we also show how postselected dual-rail systems can outperform conventional single-rail systems, accounting for qubit idling at the kiloquop scale.

\section{Background}
\label{sec:background}

\subsection{Erasure noise}

In classical information theory, an error most conventionally takes the form of a bit flip at an unknown (or `unheralded') location, where the \emph{binary symmetric channel} is a noise model which flips each bit to the opposite state with uniform probability. A linear code with distance $d$ can detect up to $d-1$ flip errors, or correct up to $\lfloor(d-1)/2\rfloor$ flip errors \cite{macwilliamsTheoryErrorcorrectingCodes1977}.\footnote{A code's \emph{distance} is the minimum Hamming distance between any two codewords, which is equivalent to the weight of the smallest undetectable error.} In contrast, an \emph{erasure} is the loss of state (i.e.\ neither 0 nor 1) at a known (or `heralded') location, where the equivalent \emph{binary erasure channel} erases each bit with uniform probability \cite{mackayInformationTheoryInference2003}. As the location is known, an appropriate code/decoder can correspondingly correct up to $d-1$ erasures.

In quantum information theory, unheralded stochastic noise generalises to a statistical ensemble of Pauli operators acting on the logical subspace:
\begin{align}
    \quantumop{P}(\rho) = &(1-p_X-p_Y-p_Z)\rho \\
    &+ p_X X\rho X + p_Y Y\rho Y + p_Z Z\rho Z \ . \notag
\end{align}
A special case is the depolarising channel, wherein the probability associated with each Pauli operator is equal, such that the state $\rho$ is mapped towards the maximally-mixed state \cite{nielsenQuantumComputationQuantum2010}:
\begin{align}
    \quantumop{D}(\rho) &= (1-p)\rho + \frac{p}{3}(X\rho X + Y\rho Y + Z\rho Z) \\
    &= (1-p)\rho + p\frac{I}{2} \ .
\end{align}

Correspondingly, an erasure is distinguished from Pauli noise by the fact that the location (and, often, the time) of the error is known; rather than requiring inference by decoding stabiliser measurements, erasure events provide classical information indicating that a qubit has left the computational subspace. Erasure-dominance can thus be a desirable system property, with highly-efficient codes \cite{reedPolynomialCodes1960, grasslCodesQuantumErasure1997} and decoders \cite{delfosseLineartimeMaximumLikelihood2020, griffithsUnionfindQuantumDecoding2024} designed for this noise channel. This has been considered an advantage of quantum architectures with natural erasure-dominance, such as dual-rail optical qubits, wherein photon loss is a major error mechanism \cite{kokLinearOpticalQuantum2007}.

\subsection{Modelling dual-rail qubits}

In conventional superconducting devices such as transmons, the primary error mechanisms are time-varying amplitude damping and phase damping, characterised by time constants $T_1$ and $T_2$ respectively, such that depolarisation is a dominant noise channel \cite{nielsenQuantumComputationQuantum2010, burnettDecoherenceBenchmarkingSuperconducting2019}. Contrastingly, leakage into non-computational states is less statistically significant. \emph{Erasure qubits} are instead engineered such that leakage events are both dominant and heralded \cite{kubicaErasureQubitsOvercoming2023}. For example, transmons can be used to construct the dual-rail encoded \emph{dimon} qubit (DDQ) \cite{willsErrordetectedCoherenceMetrology2025}; as a direct analogue of a dual-rail photonic qubit, it encodes one logical qubit into two physical modes $\ket{ab}$, such that the computational basis can be defined as
\begin{equation}
    \ket{0_L} = \ket{10} \ , \ \ket{1_L} = \ket{01} \ ,
\end{equation}
where $\ket{10}$ denotes a single excitation in mode $a$ with none in mode $b$, and vice versa. The logical codespace is therefore the single-excitation manifold
\begin{equation}
    C = \vecspan\{\ket{10},\ket{01}\} \ .
\end{equation}

A key feature of this encoding is that thermal (i.e.\ spin--lattice $T_1$) relaxation corresponds to loss of excitation, such that the energy transitions
\begin{equation}
    \ket{10}\mapsto\ket{00} \ , \ \ket{01}\mapsto\ket{00} \ ,
\end{equation}
take the system out of the logical subspace $C$. Thus, any decay into the noncomputational ground state $\ket{e}=\ket{00}$ is an erasure event. This can be interpreted as a physical error-detecting code for the erasure channel.\footnote{Note that this is \emph{not} a linear code, as $C$ is not closed under addition (readily apparent from the fact that the all-zero state is not a codeword).}

We can therefore define the erasure subspace as
\begin{equation}
    E = \vecspan\{\ket{e}\} = \vecspan\{\ket{00}\} \ ,
\end{equation}
and note that thermal relaxation becomes a detectable leakage process $C \to E$.

This erasure noise channel may similarly be written explicitly in an operator--sum form as
\begin{align}
    \quantumop{E}(\rho) &= (1-p_e) \quantumop{E}_0 \rho \quantumop{E}_0^\dag + p_e \quantumop{E}_1 \rho \quantumop{E}_1^\dag \ , \\
    \quantumop{E}_0 &= I \ , \\
    \quantumop{E}_1 &= \ket{e}\bra{10}+\ket{e}\bra{01} \ ,
\end{align}
where the erasure rate $p_e$ is the uniform probability with which each dual-rail qubit transitions into the erased state $\ket{e}$. Conditioned on no erasure, residual noise may still act as an effective Pauli channel within the logical codespace $C$, such that a convenient overall phenomenological noise model takes the form
\begin{equation}
    \quantumop{N}(\rho) = (1-p_e) \quantumop{P}(\rho) + p_e \quantumop{E}_1 \rho \quantumop{E}_1^\dag \ ,
\end{equation}
capturing the physically-relevant regime in which erasures dominate but a background of Pauli-type errors remain.

A crucial distinction between erasure noise and Pauli noise is that erasures induce conditional circuit evolution; once a qubit has left the computational subspace, subsequent gates no longer correspond to meaningful logical operations. Take $U_L$ to be any unitary matrix representing a coherent operation within the logical subspace $C$ (i.e.\ a quantum circuit gate) such that
\begin{align}
    &U_L : C \to C \ , \\
    \text{i.e.} \ &U_L\ket{\psi} = \ket{\phi} \ , \ \ket{\phi} \in C \quad \forall \ket{\psi} \in C \ .
\end{align}
Then, in our model, we treat the dynamics of the system after an erasure out of $C$ as
\begin{equation}
    U_L\ket{e} = \ket{e}
\end{equation}
such that the erased state $\ket{e}$ is stabilised by all logical operators $U_L$; that is, all subsequent operations act trivially on the erased state. Operationally, this corresponds to replacing all circuit gates on an erased qubit with no-ops.

More generally, for an erasure qubit with arbitrarily-sized $E$,
\begin{equation}
    U_L : E \to E \ ,
\end{equation}
such that the orthogonal subspaces $C$ and $E$ are disjoint under logical operators.

This model captures the physical intuition that once the encoded excitation is lost, the logical information is irrecoverable -- it cannot be restored by the remainder of the circuit. Importantly, this conditional structure must be handled carefully when performing circuit-level simulations or constructing detector--error models (DEMs) \cite{derksDesigningFaulttolerantCircuits2025}.

\subsection{Postselection}

\emph{Postselection} is the reduction of a probability distribution conditioned on the outcome of some event(s); as a hypothetical computational feature, it has been shown to significantly improve the complexity space of nondeterministic algorithms -- both classical and quantum -- by conditioning the outcome ensemble upon successful or otherwise-desired states. More formally, the classical bounded-error probabilistic polynomial-time complexity class $\complexity{BPP}$ is improved to $\complexity{BPP_{path}}$ with postselection \cite{hanThresholdComputationCryptographic1997}, and the equivalent quantum class $\complexity{BQP}$ improved to $\complexity{PostBQP}$ (which is equal to the large $\complexity{PP}$) \cite{aaronsonQuantumComputingPostselection2005}.

In this work, we investigate the practical application of this conceptual technique as error mitigation for the erasure channel, in which we simply reject circuit shots for which any erasure events were detected.

Let us consider the experimentally-motivated setting where erasure detection is performed exactly once at the end of circuit execution, a.k.a.\ \emph{end-of-line} (EOL). In the ideal limit of perfect erasure detection, postselection removes all $T_1$ events, yielding behaviour approaching an infinite-$T_1$ device, up to re-excitations. A primary objective of this paper is to quantify how closely practical dual-rail systems can approach this limit.

In realistic devices, erasure detection is imperfect. We model errors on the end-of-line erasure checks (one bit per logical qubit) as a binary asymmetric channel with the following parameters:
\begin{enumerate}
    \item false positive ($0 \mapsto 1$) rate $p_{\mathrm{fp}}$;
    \item false negative ($1 \mapsto 0$) rate $p_{\mathrm{fn}}$.
\end{enumerate}
False positives cause successful runs to be rejected, reducing sampling efficiency, whilst false negatives cause unsuccessful runs to be allowed into the accepted ensemble, limiting the overall improvement in logical fidelity. These effects therefore define a trade-off between improved error suppression and sampling overhead, with the utility of postselection depending critically on both the physical erasure rate and the fidelity of erasure checks.

The remainder of this paper develops quantitative models of this improvement, explores the role of imperfect erasure detection, and compares the resulting performance against the infinite-$T_1$ limit.

\section{Numerical techniques}
\label{sec:numerical-techniques}

\begin{figure*}
    \centering

    \begin{subfigure}{\linewidth}
        \includegraphics[width=\linewidth]{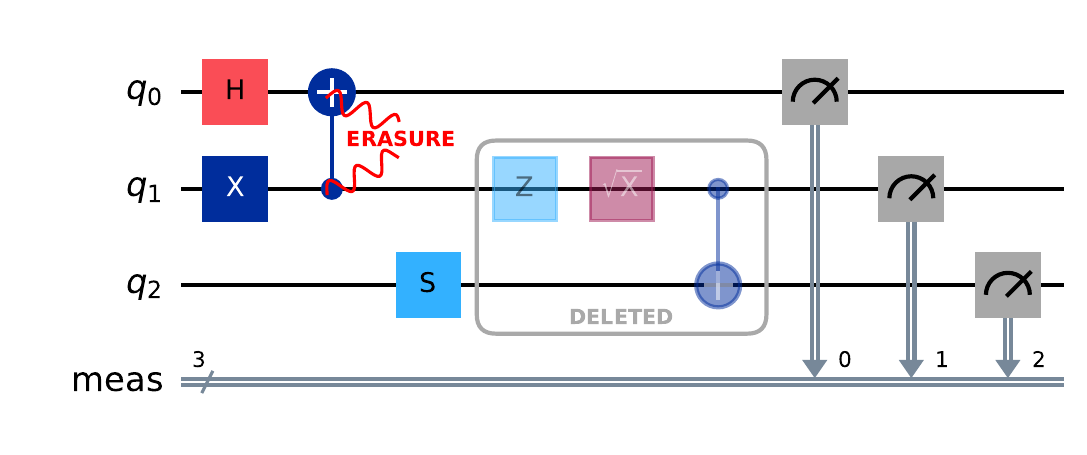}
        \caption{Example shot with the erasure circuit sampler. An erasure is stochastically induced on $q_0$ and $q_1$ by the first $\gate{CX}$ gate, so all subsequent gates touching these qubits are deleted from the circuit representation.}
        \label{subfig:circuit-sampler}
    \end{subfigure}
    \begin{subfigure}{\linewidth}
        \includegraphics[width=0.95\linewidth, trim={0 0 18.4cm 0}, clip, valign=m]{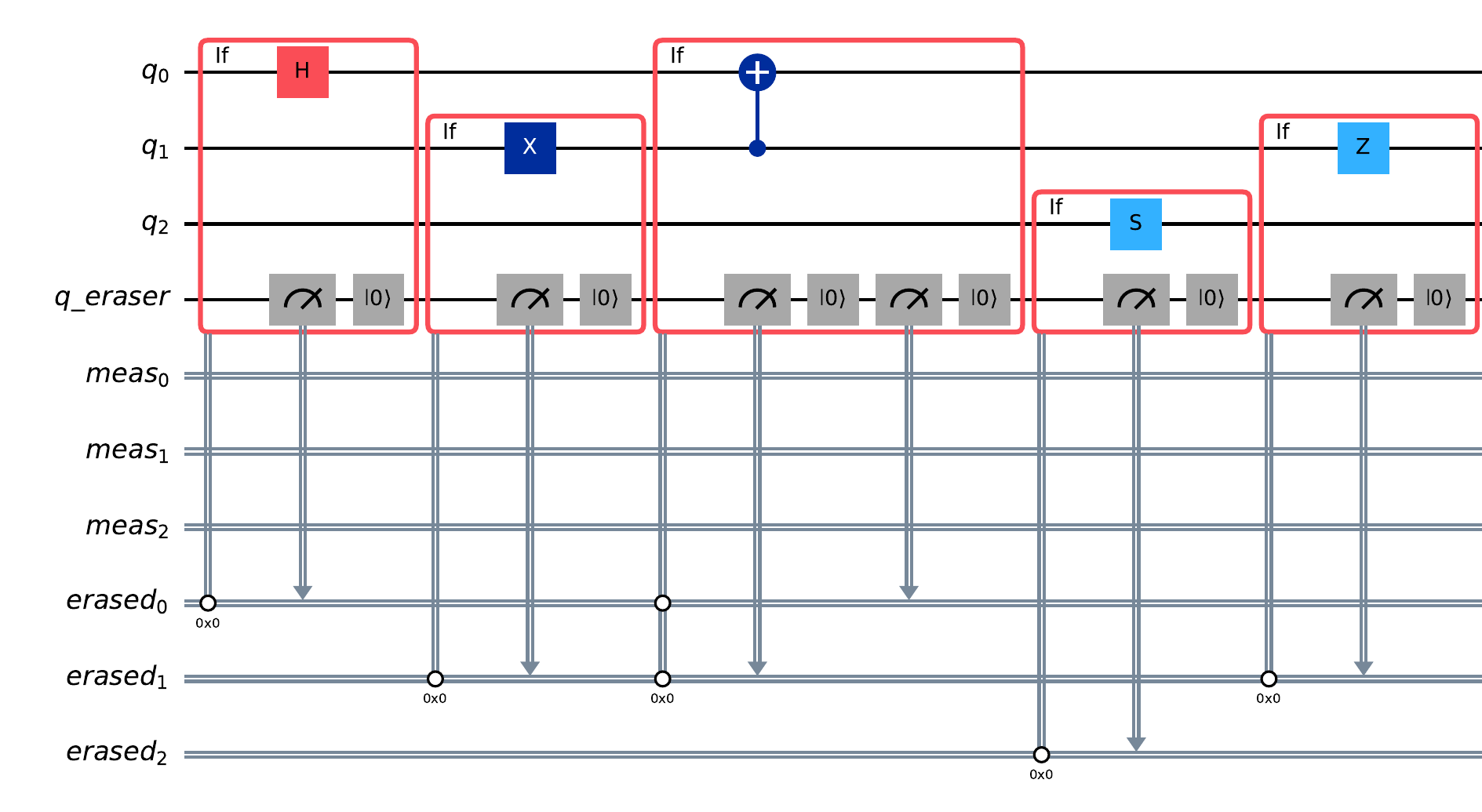} ...
        \caption{The same base circuit (cropped for clarity) instead transformed by the erasure transpiler pass. An additional classical register stores an erasure flag for each data qubit. An auxiliary qubit $q_\textrm{eraser}$ populates erasures independently onto these wires via an auxiliary bit-flip channel with rate $p_e$ on its measurement operations. All gates on data qubits are then conditioned on all qubit arguments having unset erasure flags.}
        \label{subfig:circuit-pass}
    \end{subfigure}
    
    \caption{Illustrations of the two equivalent erasure noise models implemented in \erado{} with a toy circuit example.}
    \label{fig:circuit}
\end{figure*}

\subsection{Circuit sampling/transpilation}

In this section, we outline the techniques used in our open-source erasure--postselection codebase \erado{}, which implements a circuit-level noise simulation capable of applying heralded erasure dynamics to arbitrary quantum circuits.

Two equivalent implementations of this model are provided, both of which are illustrated in Figure~\ref{fig:circuit}. The first of these (the \texttt{ErasureCircuitSampler}) enforces erasure noise by deleting gates affected by erasure events from the circuit (Figure~\ref{subfig:circuit-sampler}). This approach directly models the conditional circuit evolution induced by loss: once a qubit is erased, subsequent operations acting on that qubit no longer implement meaningful dynamics within the computational subspace and are therefore treated as no-ops. Rather than inserting additional operations or markers into the circuit, the sampler generates a modified circuit for each shot by removing the gates that would act on erased degrees of freedom. The principal benefit is that no additional overhead is introduced into the circuit representation itself; the principal drawback is that circuits are sampled and simulated on a per-shot basis.

The second of these (the \texttt{ErasurePass}) instead implements a circuit transpilation pass which adds a classical register containing the Boolean erasure state of each qubit, and wraps all gates in classical logic such that they only execute conditionally on none of the qubit arguments being erased (Figure~\ref{subfig:circuit-pass}). The principal benefit is that this yields a single precomputed circuit which fully implements the noise model across all shots; the principal drawback is that the addition of classical logic significantly slows down the simulation of quantum circuits.

With either implementation, this model occupies a complementary regime to simulational techniques based on stabilisers and detector--error models (DEMs) \cite{gidneyStimFastStabilizer2021}. It captures conditional dynamics intrinsic to erasure processes and supports arbitrary Qiskit circuits (and backends, in the case of the circuit sampler), at the cost of either per-shot circuit sampling or the introduction of classical control flow. This makes it well-suited to near-term studies of erasure-aware primitives, postselection strategies, and the impact of imperfect erasure checks; meanwhile, scalable DEM-based pipelines remain essential for large-distance QEC studies and broad architectural exploration.

In the remainder of this paper, of these two implementations, we refer to and use the \texttt{ErasureCircuitSampler} for numerically characterising erasure-induced behaviour in circuits (empirically, it outperformed the erasure transpiler pass in overall runtime on our high-performance computing platforms).

\subsection{Sampling semantics}

We assume a uniform erasure rate $p_e$ applied across the circuit execution. The circuit sampler supports two closely-related semantics controlled by a Boolean option \texttt{erasure\_before\_gates}:
\begin{enumerate}
    \item \emph{post-gate erasure}, where erasure is inflicted immediately after a gate acts;
    \item \emph{pre-gate erasure}, where erasure is inflicted immediately before a gate acts.
\end{enumerate}
These two conventions represent common physical interpretations (e.g.\ erasure associated with idle evolution, versus erasure associated with driven operations) and allow sensitivity studies to bracket hardware-dependent timing assumptions. In both cases, once an erasure event occurs on a qubit, all subsequent gates that touch that qubit are removed from the shot circuit. For multi-qubit gates, this deletion rule naturally removes any operation for which one participant has been erased.

Take a circuit represented as an ordered collection of gates $G$ which, as a tuple, can be defined as a set of index--operator pairs $(t,U)$ in the form
\begin{equation}
    G = \{ (1,U_1) , (2,U_2) , \dots , (\abs{G},U_{\abs{G}}) \} \ .
\end{equation}
We denote the gate index with $t$ as it is hereinafter used interchangeably with the more general concept of time, under our chosen circuit timing convention. Each element in $G$ is therefore a candidate location for an erasure event. For each shot $\sigma$, the sampler stochastically draws a vector of independent erasure flags
\begin{equation}
    \vec{e}_\sigma = \left( {e_\sigma}_t \mid t \in [1 \twodots \abs{G}], {e_\sigma}_t \sim \mathrm{Bernoulli}(p_e) \right) \ ,
\end{equation}
yielding a set of erasure locations for each shot
\begin{equation}
    \Omega_\sigma = \left\{ t \mid {e_\sigma}_t = 1 \right\} \ .
\end{equation}

\subsection{Deletion lookup table (LUT)}

To avoid computing deletion sets repeatedly over many shots, the \texttt{ErasureCircuitSampler} precomputes a lookup table (LUT) at construction time. Given an input circuit, we enumerate the ordered gate representation $G$ and, treating each element as a candidate erasure location, cache the set of downstream gates which would be deleted.

Formally, each erasure event at location $t$ maps to a deletion set
\begin{equation}
    \tau(t) \subseteq [t \twodots \abs{G}] \ ,
\end{equation}
consisting of all gate indices $t' \ge t$ where the support of $U_{t'}$ includes any of the qubits in the support of the deleted gate $U_t$ (if using post-gate erasure, the above subset and inequality become strict).

The lookup table therefore enables $O(1)$ retrieval of the relevant deletion set for each sampled erasure event, reducing the problem of shot generation to set union and filtering. This design is particularly effective for parameter sweeps, since the lookup table depends only on circuit structure and erasure timing convention, and thus can be reused across many runs with different erasure rates, backend configurations, or analysis modes.

\subsection{Shot generation and execution}

For each shot $\sigma$, an erasure-sampled circuit $G_\sigma$ is constructed by deleting from $G$ the union of deletion sets implied by the erasure locations $\Omega_\sigma$; that is, the tuple of \emph{all} deleted gate locations $D_\sigma$ is
\begin{equation}
    D_\sigma = \left\{ (t', U_{t'}) \middlemid t' \in \bigcup_{t\in\Omega_\sigma} \tau(t) \right\} \ ,
\end{equation}
such that
\begin{equation}
    G_\sigma = G \setminus D_\sigma \ .
\end{equation}

The \texttt{ErasureCircuitSampler} implementing erasure noise as gate deletion means that the resulting shot circuit $G_\sigma$ contains only a subset of the original circuit primitives, introducing no additional operations, measurements or classical control flow (in contrast to the \texttt{ErasurePass}). In this sense, the method enforces conditional `no-op' dynamics without modifying the gate set.

\subsection{Randomness and concurrency}

The \texttt{ErasureCircuitSampler} implements reproducible stochastic sampling through a multiprocessing-safe random number generator (RNG) interface (\texttt{MultiprocessingRNG}). Importantly, the \texttt{seed} method controls the entropy used to generate erasure patterns, and also correctly overrides the seed behaviour of the backend used for simulating circuit shots. This ensures that both erasure sampling and circuit execution are reproducible and stable under parallel execution, enabling deterministic reruns and statistically-valid comparisons across parameter sweeps.

Because shots are independent, this method is embarrassingly parallel; in practice, we distribute shots across worker processes. By aggregating desired results\footnote{(e.g.\ accepted/rejected proportions, logical state measurements and circuit fidelities)} via multiprocessing-safe optional callbacks, we also avoid the need to store full shot-level traces (unless required for debugging or other detailed analysis).

Finally, since erasure-aware simulation can occasionally produce pathological cases for certain backends or compilation settings (e.g.\ extreme deletion patterns, backend-specific slowdowns or system-level multiprocessing limits), the \texttt{ErasureCircuitSampler} supports a per-shot \texttt{timeout} parameter, which upper-bounds the allowed execution time for a single shot before termination. This provides robust behaviour in large Monte Carlo simulations and prevents a small number of outliers from dominating total runtime.

\section{Results}
\label{sec:results}

\subsection{Model and metrics}


\begin{figure}
    \begin{subfigure}{\linewidth}
        \includegraphics[width=\linewidth]{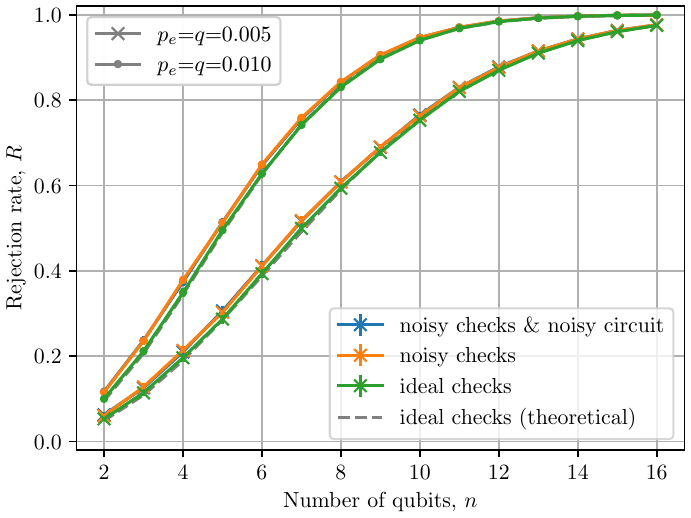}
        \caption{}
        \label{subfig:rejection-rate}
    \end{subfigure}
    \begin{subfigure}{\linewidth}
        \includegraphics[width=\linewidth]{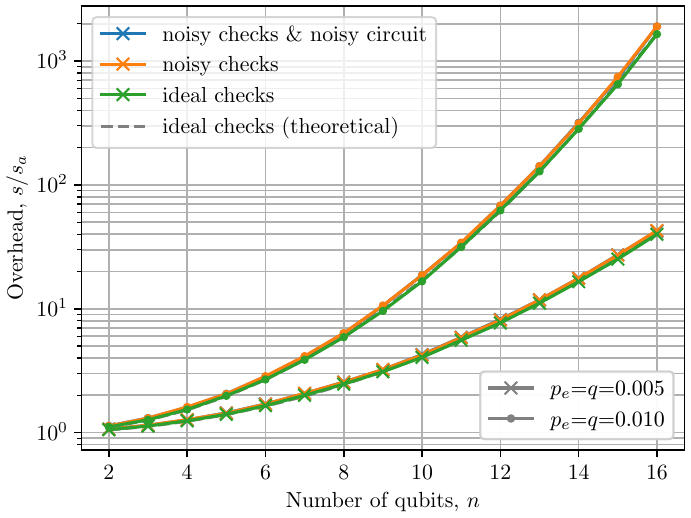}
        \caption{}
        \label{subfig:overhead}
    \end{subfigure}

    \caption{Cost of postselection for QFT with 5000 target shots. The trends show erasure noise with ideal checks (green), erasure noise with check noise (orange), and erasure noise with check noise and gate depolarising noise (blue). Note how false positives dominate to strictly increase the rejection rate over ideal checks. Error bars for rejection rate show the $95\%$ confidence interval, calculated via the Clopper-Pearson method for binomial proportions.}
    \label{fig:rejection-rate-and-overhead}
\end{figure}

In these results, we study the effects of noise channels and postselection on the performance of a linear-connectivity quantum Fourier transform (QFT) construction, as given in \cite{fowlerImplementationShorsAlgorithm2004}. Arbitrarily but consistently, all circuits are transpiled to the universal basis gate set $\{\gate{R_Z}, \gate{\sqrt{X}}, \gate{CX}\}$. For a number of qubits $n\in[2 \twodots 16]$, the number of gates $\abs{G}$ in this QFT construction scales quadratically as
\begin{equation} \label{eq:qft-gates}
    \abs{G} = 3n^2-2n+2 \ .
\end{equation}

The erasure rate $p_e$ is the uniform probability with which any of the $\abs{G}$ gates in the circuit inflicts an erasure upon any of its qubit arguments. This is equivalent to an amplitude damping channel with emission rate $p_e$, where $p_e$ models thermal relaxation by assuming the time-varying form $1-e^{-t/T_1}$ \cite{nielsenQuantumComputationQuantum2010}. For example, a gate time $t=\qty{500}{\nano\second}$ and a physical relaxation time $T_1=\qty{50}{\micro\second}$ equates to an erasure rate $p_e \approx 0.010$.

Erasure check noise is modelled with a single error rate $q$ for simplicity, encompassing both false positives and false negatives as $q=p_\mathrm{fp}=p_\mathrm{fn}$. Unheralded Pauli noise is modelled as a depolarising channel with 1-qubit gate error rate $p_\mathrm{1Q}=0.001$ and 2-qubit rate $p_\mathrm{2Q}=0.010$.

When postselection is enabled, the simulation runs until the number of accepted shots $s_a$ equals some target (e.g.\ 5000 shots). This leads to a number of rejected shots $s_r$, with total shots $s=s_a+s_r$. The \emph{rejection rate} $R$ is thus defined as the proportion of rejected shots:
\begin{equation} \label{eq:rejection-rate}
    R = \frac{s_r}{s} = 1-\frac{s_a}{s} \ .
\end{equation}
A shot is rejected if any qubits are erased, such that the theoretical rejection rate scales binomially as
\begin{equation} \label{eq:rejection-rate-trend}
    R = 1-(1-p_e)^{\abs{G}} \ .
\end{equation}
Correspondingly, we define the \emph{overhead} of postselection as the proportion of total shots to accepted shots $s/s_a$. The theoretical trend for the overhead is derived by combining Equations \ref{eq:rejection-rate} and \ref{eq:rejection-rate-trend}:
\begin{align} \label{eq:overhead}
    \frac{s}{s_a} &= (1-p_e)^{-\abs{G}} \\
    &\in (1-p_e)^{-O(n^2)} \ .
\end{align}
Therefore, we can say that this postselection problem scales exponentially, in that its overhead scales with a polynomial exponent in $n$ (as per the complexity class $\complexity{EXPTIME}$). We verify these theoretical trends for the rejection rate and overhead empirically in Figure~\ref{fig:rejection-rate-and-overhead}.

Finally, the per-shot circuit fidelity $F$ is defined conventionally as
\begin{equation}
    F(\ket{\psi}, \ket{\Psi}) = \abs{\braket{\psi | \Psi}}^2 \ ,
\end{equation}
where $\ket{\psi}$ is the end-of-line statevector snapshot of the simulation (before final measurements) and $\ket{\Psi}$ is the ideal statevector representation calculated directly from the circuit definition for each $n$.

\begin{figure}
    \begin{subfigure}{\linewidth}
        \includegraphics[width=\linewidth]{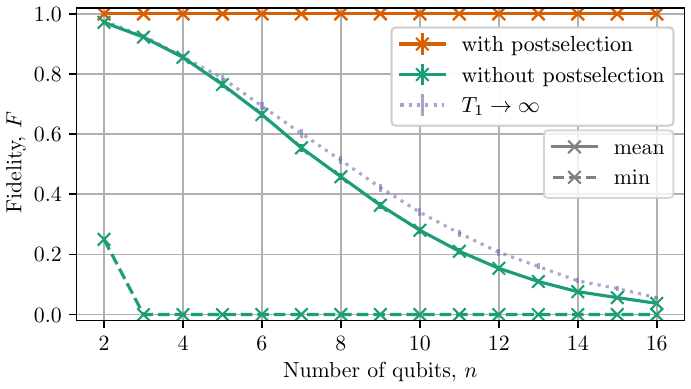}
        \caption{Ideal checks and ideal circuit. Note how postselection definitionally yields $F=1.0$.}
        \label{subfig:fidelity-idealchecks-idealcirc}
    \end{subfigure}
    \begin{subfigure}{\linewidth}
        \includegraphics[width=\linewidth]{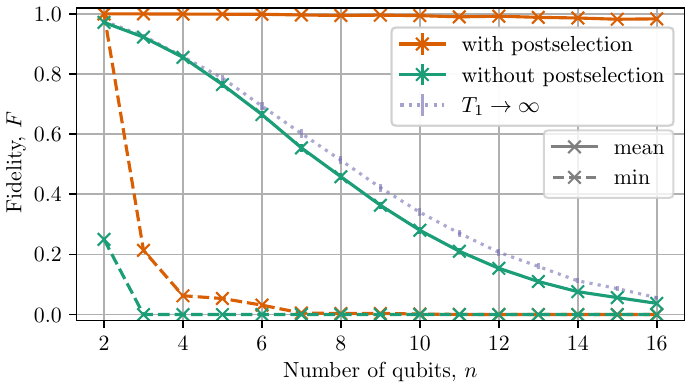}
        \caption{Noisy checks and ideal circuit. Note the small-but-nonzero effect of false negatives upon mean fidelity.}
        \label{subfig:fidelity-noisychecks-idealcirc}
    \end{subfigure}
    \begin{subfigure}{\linewidth}
        \includegraphics[width=\linewidth]{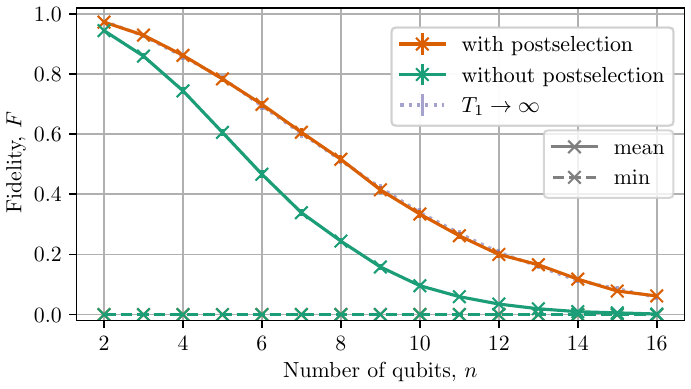}
        \caption{Noisy checks and noisy (depolarising) circuit. Note how postselection approximately achieves $T_1\to\infty$, excepting false negatives.}
        \label{subfig:fidelity-noisychecks-noisycirc}
    \end{subfigure}

    \caption{Mean and minimum circuit fidelity for QFT with 5000 target shots, with and without postselection, for $p_e=0.005$ and $q=0.010$. The $T_1\to\infty$ reference is equivalent to a standalone depolarising model (i.e. $p_e=q=0$). Error bars for mean fidelity (here and hereinafter) show the $95\%$ confidence interval, calculated via Student's $t$ distribution of the standard error of the mean (SEM).}
    \label{fig:fidelity}
\end{figure}

\subsection{Interaction of noise models}
The impact of postselection on the fidelity $F$ of the QFT algorithm is presented in Figure~\ref{fig:fidelity}; in particular, we observe how erasure and postselection impacts the fidelity with respect to the limit of $T_1\to\infty$, i.e.\ the regime where $p_e=0$ and the remaining unheralded Pauli errors (i.e.\ our gate depolarising channel) alone constitute the noise model. The fidelity for this regime is included across all three subfigures -- note that this is a deterministic reference series, calculable directly via a density-matrix representation of the circuit:
\begin{align}
    \ket{\Psi} &= (U_{\abs{G}} \dots U_2U_1)\ket{0}^{\otimes n} \ , \\
    \rho &= \ket{\Psi}\bra{\Psi} \ , \\
    F(\rho,\quantumop{D}(\rho)) &= \left( \trace{\sqrt{\sqrt{\rho} \quantumop{D}(\rho) \sqrt{\rho}}} \right)^2 \ . \label{eq:fidelity-dm}
\end{align}

Firstly, Figure~\ref{subfig:fidelity-idealchecks-idealcirc} shows the impact of postselection on QFT with $p_e=0.005$ perfect erasure checks ($q=0$) and no depolarising noise. With perfect postselection, the mean (and minimum) fidelity is by definition improved to exactly 1.0, because erasures constitute the sole noise channel, which is perfectly mitigated.

Secondly, Figure~\ref{subfig:fidelity-noisychecks-idealcirc} shows the same simulation but with imperfect erasure checks, where $q=p_\mathrm{fp}=p_\mathrm{fn}=0.010$. Here, we can see how false negatives have a distinct but small impact on the expected $F=1.0$ result with postselection, by erroneously allowing shots experiencing erasure into the accepted ensemble. Note again that in contrast, false positives do not affect the fidelity, but instead drive up the rejection rate/overhead (as is demonstrated starkly in Figure~\ref{fig:rejection-rate-and-overhead}).

Thirdly, Figure~\ref{subfig:fidelity-noisychecks-noisycirc} shows the same simulation but with imperfect erasure checks \emph{and} unheralded gate depolarising noise ($p_\mathrm{1Q}=0.001,p_\mathrm{2Q}=0.010$). Two key points are immediately apparent from this data:
\begin{enumerate}
    \item without postselection, the erasure and depolarising channels have stacking negative effects on fidelity;
    \item with postselection, the erasure channel is mitigated such that the mean fidelity accurately matches $\lim_{T_1\to\infty}F$, excepting false negatives.
\end{enumerate}

\begin{figure}
    \begin{subfigure}{\linewidth}
        \includegraphics[width=\linewidth]{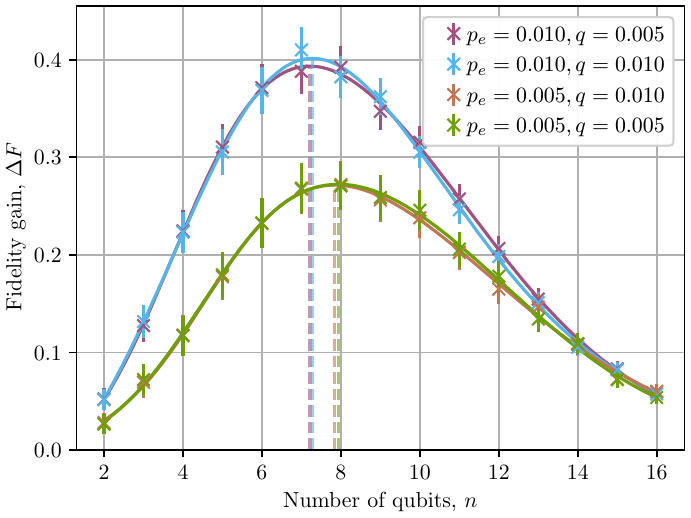}
    \end{subfigure}
        \begin{subfigure}{\linewidth}
        \includegraphics[width=\linewidth]{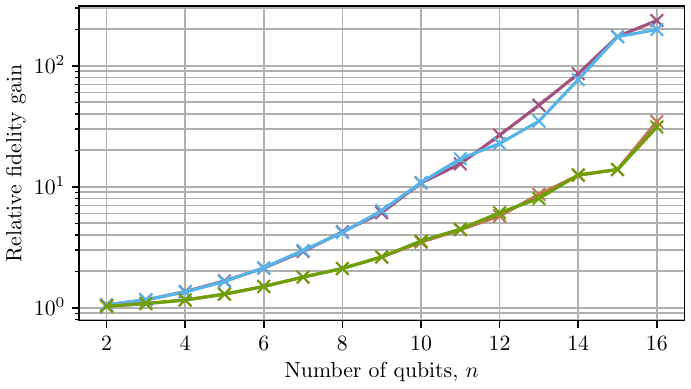}
    \end{subfigure}
    \caption{Gain in mean circuit fidelity for QFT achieved with postselection. For the absolute fidelity gain $\Delta F$ (top), the curves are fitted via nonlinear least-squares regression on the difference between two reverse logistic (i.e.\ sigmoid) functions, as in Figure~\ref{subfig:fidelity-noisychecks-noisycirc} (with error bars propagated accordingly), with their approximated maxima shown. The relative gain (bottom) shows the quotient rather than the difference.}
    \label{fig:fidelity-gain}
\end{figure}

To summarise, this data shows the dynamics of adding in three different `layers' of noise channel:
\begin{enumerate}
    \item gate-deleting erasure noise;
    \item classical bit-flip noise on the erasure checks;
    \item gate depolarising noise as unheralded Pauli error.
\end{enumerate}
It is clear that postselection fully mitigates the erasure channel, such that $F=1.0$ in the absence of other channels and $F=F(\rho,\quantumop{D}(\rho))$ in the presence of a depolarising channel $\quantumop{D}$, with an apparently minor impact by false negatives when $q=0.010$.

In the next two subsections, we consider the following emergent questions: \emph{by how much} does postselection improve fidelity with the full noise model (Section~\ref{subsec:fidelity-gain}), and \emph{how close to} $\lim_{T_1\to\infty}F$ can we get with imperfect-check postselection (Section~\ref{subsec:approaching-limit})?

\subsection{Fidelity gain with postselection}
\label{subsec:fidelity-gain}
\begin{table}
    \centering
    \begin{tabular}{ |c|c| } 
         \hline
         $q$ & $\Delta F_{T_1\to\infty}$ \\
         \hline
         0.005 & $\ \: \, 0.0022 \pm 0.0095$  \\
         0.010 & $-0.0010 \pm 0.0095$ \\
         0.015 & $-0.0073 \pm 0.0095$ \\
         0.050 & $-0.0219 \pm 0.0095$ \\
         \hline
    \end{tabular}

    \caption{Drop in mean circuit fidelity for $p_e=0.010$ and $n=7$ for QFT with 10,000 target shots (as seen in Figure~\ref{fig:q-sweep}). Note that for $q < 0.030$, $\Delta F_{T_1\to\infty}$ is within error of zero.}
    \label{table:fidelity-drop}
\end{table}

Figure~\ref{fig:fidelity} shows how -- for both the erasure and depolarising channels -- the mean fidelity decays with a reverse sigmoid (i.e.\ S-shaped) curve as $n$ increases. Of particular interest is how postselection under gate depolarising noise shallows this curve compared to no postselection (Figure~\ref{subfig:fidelity-noisychecks-noisycirc}), appearing to yield a gap of maximum magnitude at exactly one point. This gap therefore corresponds to the maximum absolute gain in circuit fidelity $\Delta F$ achieved with postselection.

To study this further, we repeat the simulation for four different combinations of $p_e$ and $q$ and plot $\Delta F$ as a function of $n$ in Figure~\ref{fig:fidelity-gain}. As we may intuit from the sigmoids by inspection, $\Delta F(n)$ takes the form of a unimodal bell-shaped curve.\footnote{In fact, the integral of any unimodal bell-shaped curve is sigmoidal by definition.} For each series, we fit the curve as a difference of two reversed logistic functions and maximise this function, yielding
\begin{align}
    \argmax_n{\Delta F} &\approx 7 \ , \quad p_e=0.010 \ , \\
    \argmax_n{\Delta F} &\approx 8 \ , \quad p_e=0.005 \ .
\end{align}
In other words, as $p_e$ increases, the maximum fidelity gain from postselection also increases, but occurs for lower $n$. Equivalently, the \emph{relative} fidelity gain increases monotonically with $n$.

\begin{figure}
    \centering
    \includegraphics[width=\linewidth]{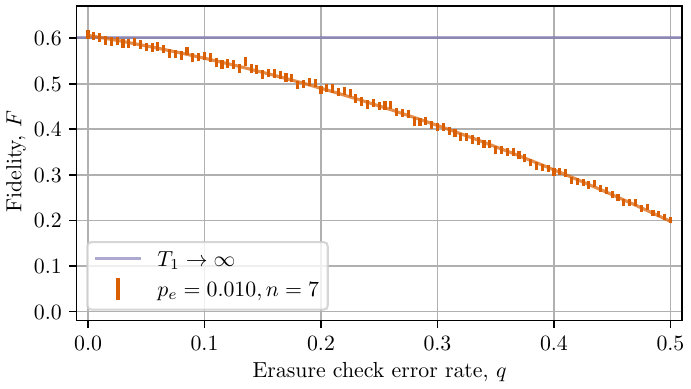}
    \caption{Mean circuit fidelity for QFT with 10,000 target shots, for $p_e=0.010$ and $n=7$. As the check error rate $q$ tends to zero, postselection perfectly mitigates the erasure channel, leaving only the gate depolarising channel. The curve fit is a `quadratic decay' of the form $F=d - (aq^2+bq+c)$, approximating the polynomial structure in the binomial expansion of false-negative events.}
    \label{fig:q-sweep}
\end{figure}

\subsection{Approaching the limit of \texorpdfstring{$T_1\to\infty$}{T1toinfinity}}
\label{subsec:approaching-limit}
Taking $p_e=0.010$ and $n=7$ as an example `sweet spot' of maximum fidelity gain from the previous section (Figure~\ref{fig:fidelity-gain}), let us now consider the question of how effectively postselection mitigates against the erasure channel given imperfect erasure checks. For these values, due to gate depolarising noise, the expected fidelity with perfect postselection (from Equation~\ref{eq:fidelity-dm}) is
\begin{equation}
    \lim_{T_1\to\infty}{F} \approx 0.6021836859864419 \ .
\end{equation}

In Figure~\ref{fig:q-sweep}, we plot the mean circuit fidelity and sweep the full entropy space of the erasure check error rate $0 \leq q \leq 0.5$, observing that
\begin{equation}
    \lim_{q \to 0}{F} = \lim_{T_1\to\infty}{F} \ .
\end{equation}
As $q$ increases, $F$ decays from $\lim_{T_1\to\infty}{F}$ by a polynomial factor arising from the binomial expansion of false-negative events.

From this data, we indicate in Table~\ref{table:fidelity-drop} the drop in mean circuit fidelity (compared to perfect postselection) caused by some realistic NISQ-era benchmarks for $q$, noting that for $q < 0.030$, $\Delta F_{T_1\to\infty}$ is within error of zero.\footnote{See Figure~\ref{fig:fidelity} for error calculation.} These values are consistent with the small influence of imperfect checks under the erasure channel seen between Figures~\ref{subfig:fidelity-idealchecks-idealcirc} and Figures~\ref{subfig:fidelity-noisychecks-idealcirc}.

\subsection{Quantum operation benchmarks}
\label{subsec:quop-benchmarks}
\begin{table}
    \centering
    \begin{tabular}{ |c|c|c|c|c| }
         \hline
         $n$ & $\abs{G}$ & $p_\mathrm{1Q}$ & $p_\mathrm{2Q}$ & $\Delta F$ \\
         \hline
         5 & 149 & 0.016 & 0.050 & 0.03330360 \\
         10 & 514 & 0.004 & 0.013 & 0.07587702 \\
         15 & 1029 & 0.002 & 0.005 & 0.14007602 \\
         \hline
    \end{tabular}

    \caption{Depolarising error budgets for QFT in the $\qty{100}{\quop}$, $\qty{500}{\quop}$ and $\qty{1}{\kilo\quop}$ regimes. The improvement in mean fidelity from single-rail to postselected dual-rail models (as seen in Figure~\ref{fig:quop-benchmarks}) is listed as $\Delta F$.}
    \label{table:quop-benchmarks}
\end{table}

The number of gates $\abs{G}$ is a key metric in describing the scale of quantum computation in practice, popularly given the unit \emph{quop} as a parallel to the classical \emph{flop} (i.e.\ \emph{quantum operations} versus \emph{floating-point operations}) \cite{preskillNISQMegaquopMachine2025}. We conclude our results with datasets representative of practical $\qty{100}{\quop}$, $\qty{500}{\quop}$ and $\qty{1}{\kilo\quop}$ regimes.

To compare the performance of postselected dual-rail qubits against conventional systems of comparable specifications, we replace the erasure channel with an amplitude damping channel (whilst keeping the gate depolarising channel constant). As this amplitude damping channel takes an emission rate equal to the erasure rate $p_e$, it models the same physical relaxation time $T_1$ as a single-rail qubit.

Additionally, we account for decoherence caused by qubit idling; this is a significant source of noise, which can be mitigated by techniques such as \emph{dynamical decoupling}, wherein circuit-preserving operations are performed on qubits during idle periods \cite{krantzQuantumEngineersGuide2019}. To model idling error in our simulations, noisy identity gates representing discrete idle operations are populated in the circuit using a modified dynamical decoupling scheduler (see Appendix~\ref{sec:idling-error}).

\begin{figure}
    \centering
    \includegraphics[width=\linewidth]{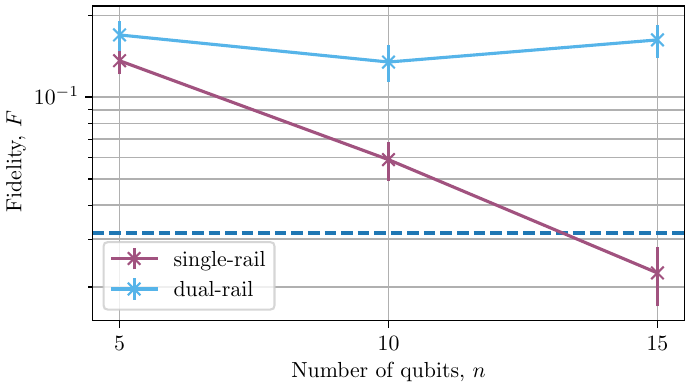}
    \caption{Mean circuit fidelity for QFT with $s_a=1000$ target shots for both single-rail and postselected dual-rail qubits, assuming $p_e=0.008$ and $q=0.005$. The dashed horizontal line marks a $1/\sqrt{1000} \approx 3\%$ fidelity benchmark \cite{hughesMeasurementsTheirUncertainties2010}. Table~\ref{table:quop-benchmarks} lists the depolarising error budgets. Idling error is included, such that the datapoints correspond to $\qty{100}{\quop}$, $\qty{500}{\quop}$ and $\qty{1}{\kilo\quop}$.}
    \label{fig:quop-benchmarks}
\end{figure}

Without idle operations, for some select $n$, the number of gates $\abs{G}(n)$ in the QFT construction as per Equation~\ref{eq:qft-gates} is
\begin{align}
    \abs{G}(5) &= 67 \ , \\
    \abs{G}(10) &= 282 \ , \\
    \abs{G}(15) &= 647 \ .
\end{align}
With idle operations added using our representation, these benchmarks increase to
\begin{align}
    \abs{G}(5) &= 149 \ , \\
    \abs{G}(10) &= 514 \ , \\
    \abs{G}(15) &= 1029 \ ,
\end{align}
thus corresponding to the $\qty{100}{\quop}$, $\qty{500}{\quop}$ and $\qty{1}{\kilo\quop}$ regimes, respectively.

Figure~\ref{fig:quop-benchmarks} compares the mean fidelity for QFT between these single-rail and postselected dual-rail models. Across both models, we assume an emission rate $p_e=0.008$ (approximately corresponding to a gate time $t=\qty{500}{\nano\second}$ and physical relaxation time $T_1=\qty{60}{\micro\second}$) with accepted shots $s_a=1000$. In the dual-rail case, we assume an erasure check error rate $q=0.005$.

The depolarising error rates $p_\mathrm{1Q}$ and $p_\mathrm{2Q}$, listed in Table~\ref{table:quop-benchmarks}, are chosen for different $n$ as rough estimates of practical error budgets, such that the factor in overall fidelity is no worse than 50\% due to 1Q gate error, and 20\% due to 2Q gate error.

We can see in Figure~\ref{fig:quop-benchmarks} that transitioning from the single-rail to the postselected dual-rail model consistently increases the mean fidelity, with this increase $\Delta F$ listed in Table~\ref{table:quop-benchmarks}. In particular, we take a fidelity benchmark of $1/\sqrt{s_a} \approx 0.03$, as an estimate of the noise floor resulting from shot noise as a Poisson process \cite{hughesMeasurementsTheirUncertainties2010}. For $n=15$, the single-rail model falls short of this minimum-required fidelity at $F\approx0.023$, but the postselected dual-rail model comfortably surpasses it at $F\approx0.163$.

\section{Conclusion}

In this work, we have presented a theoretical and simulational framework for the use of postselection with erasure-dominant qubits, e.g.\ dual-rail transmons. Acknowledging that the overhead of postselection scales exponentially due to the rejection rate (Equation~\ref{eq:overhead}; Figure~\ref{fig:rejection-rate-and-overhead}), we have demonstrated empirically how postselection can accurately mitigate the erasure channel in terms of mean circuit fidelity for QFT, even in the presence of noisy erasure checks (Figure~\ref{fig:fidelity}).

We have shown how the gain in fidelity achieved by postselection on such dual-rail systems increases with the erasure rate $p_e$ (or, equivalently, decreases against the physical relaxation time $T_1$), with the maximum gain emerging at smaller circuit sizes as $p_e$ increases (Figure~\ref{fig:fidelity-gain}). We have also shown how the fidelity converges polynomially to the limit of $T_1\to\infty$ (in which only the gate depolarising channel remains), with the difference from this limit arriving within error of zero for erasure check error rate $q<0.030$ (Figure~\ref{fig:q-sweep}; Table~\ref{table:fidelity-drop}).

Finally, we have shown that a conventional single-rail system does not surpass an estimate of the fundamental noise floor for QFT in the kiloquop regime, whereas a comparable postselected dual-rail system does (Figure~\ref{fig:quop-benchmarks}; Table~\ref{table:quop-benchmarks}). We believe these results may justify the use of erasure-based postselection as a near-term strategy for error mitigation, before (and, perhaps, combined with) the practical arrival of scalable quantum fault-tolerance.

\section*{Acknowledgements}
This work was supported by the Innovate UK Quantum Missions pilot competition 10148061 DECIDE: Dimon error correction integrated into a data-centre environment.

We also thank Bryn Bell and Ailsa Keyser for their contributions and reviews. 

\appendix
\section{Modelling idling error}
\label{sec:idling-error}

Dynamical decoupling (DD) is a technique in which a sequence of quantum operations (e.g.\ $\gate{X}\gate{X})$ that composes to the identity -- and thus does not alter the overall circuit -- is inserted into idle periods and acts to mitigate decoherence caused by qubit idling \cite{krantzQuantumEngineersGuide2019}. A DD scheduler is provided as a circuit transpiler pass by the \texttt{qiskit-ibm-runtime} package \cite{javadi-abhariQuantumComputingQiskit2024}.

To simulate idling error in our simulations in Section~\ref{subsec:quop-benchmarks}, we use this transpiler pass to insert identity gates on qubits during idle periods; these identity gates are subject to all of the same noise channels we consider as the circuit gates.

This circuit transformation is implemented in \erado{} by first performing an as-late-as-possible (ALAP) scheduling pass, and then performing a DD scheduling pass. We set every circuit gate to be exactly 1 unit of time in duration, and every identity gate inserted by the DD scheduler to be 0.8 units of time, with \texttt{sequence\_min\_length\_ratios} equal to 1.0. With a DD sequence of $\gate{I}\gate{I}$, the scheduler would therefore insert nothing into an idle period of 1 unit, and an $\gate{I}\gate{I}$ sequence into an idle period of 2 units or greater.

In order to fill arbitrarily-long idle periods with arbitrarily-many identity operators, we populate the list of possible DD sequences with even-weight sequences of repeated $\gate{I}$ gates in descending priority. Given a maximum idle sequence length $k$, where $k$ is necessarily even, the ordered list of possible idle sequences is
\begin{equation}
    ( \gate{I}^k , \gate{I}^{k-2} , \dots , \gate{I}^2 ) \ .
\end{equation}
Resultingly, the number of identity gates inserted into an idle period of length $t$ units of time is equal to
\begin{equation}
    \min\left\{ t\left\lfloor\frac{t}{2}\right\rfloor , k \right\} \ .
\end{equation}

In a relatively sparse circuit such as QFT, increasing the number of qubits $n$ notably increases the total amount of time spent idling by each qubit. We therefore use $k$ as a parameter to tune the density of idling gates populated into the circuit representation for increasingly large $n$. For example, a value of $k=14$ is used in Section~\ref{subsec:quop-benchmarks} to most closely map $\abs{G}$ to the three operational regimes discussed, which also broadly agrees with ParityQC QFT compilations \cite{klaverSWAPlessImplementationQuantum2026}. This is justified from the perspective of DD as a technique, wherein the mitigation achieved by aggressively populating DD sequences forms a trade-off with infidelity in the sequence gates themselves.

\printbibliography

@book{nielsenQuantumComputationQuantum2010,
  title = {Quantum Computation and Quantum Information},
  author = {Nielsen, Michael A. and Chuang, Isaac L.},
  date = {2010-12-09},
  edition = {10th Anniversary Edition},
  publisher = {Cambridge University Press},
  doi = {10.1017/CBO9780511976667},
  url = {https://aapt.scitation.org/doi/10.1119/1.1463744},
  urldate = {2019-01-16},
  isbn = {978-1-107-00217-3},
}

@article{fowlerImplementationShorsAlgorithm2004,
  title = {Implementation of {{Shor}}'s Algorithm on a Linear Nearest Neighbour Qubit Array},
  author = {Fowler, A. G. and Devitt, S. J. and Hollenberg, L. C. L.},
  date = {2004-07-01},
  journaltitle = {Quantum Information \& Computation},
  volume = {4},
  number = {4},
  eprint = {quant-ph/0402196},
  eprinttype = {arXiv},
  pages = {237--251},
  issn = {1533-7146}
}

@article{burnettDecoherenceBenchmarkingSuperconducting2019,
  title = {Decoherence Benchmarking of Superconducting Qubits},
  author = {Burnett, Jonathan J. and Bengtsson, Andreas and Scigliuzzo, Marco and Niepce, David and Kudra, Marina and Delsing, Per and Bylander, Jonas},
  date = {2019-06-26},
  journaltitle = {npj Quantum Information},
  shortjournal = {npj Quantum Inf},
  volume = {5},
  number = {1},
  pages = {54},
  publisher = {Nature Publishing Group},
  issn = {2056-6387},
  doi = {10.1038/s41534-019-0168-5},
  url = {https://www.nature.com/articles/s41534-019-0168-5},
  urldate = {2026-02-24},
  langid = {english},
}

@book{macwilliamsTheoryErrorcorrectingCodes1977,
  title = {The Theory of Error-Correcting Codes},
  author = {MacWilliams, Florence Jessie and Sloane, Neil James Alexander},
  date = {1977},
  eprint = {nv6WCJgcjxcC},
  eprinttype = {googlebooks},
  publisher = {Elsevier},
  isbn = {978-0-444-85010-2},
  langid = {english},
  pagetotal = {788}
}

@article{dennisTopologicalQuantumMemory2002,
  title = {Topological Quantum Memory},
  author = {Dennis, Eric and Kitaev, Alexei and Landahl, Andrew and Preskill, John},
  date = {2002-09},
  journaltitle = {Journal of Mathematical Physics},
  shortjournal = {Journal of Mathematical Physics},
  volume = {43},
  number = {9},
  eprint = {quant-ph/0110143},
  eprinttype = {arXiv},
  pages = {4452--4505},
  issn = {0022-2488, 1089-7658},
  doi = {10.1063/1.1499754},
  url = {http://arxiv.org/abs/quant-ph/0110143},
  urldate = {2020-05-30},
}

@article{calderbankGoodQuantumErrorcorrecting1996,
  title = {Good Quantum Error-Correcting Codes Exist},
  author = {Calderbank, A. R. and Shor, Peter W.},
  date = {1996-08-01},
  journaltitle = {Physical Review A},
  shortjournal = {Phys. Rev. A},
  volume = {54},
  number = {2},
  eprint = {quant-ph/9512032},
  eprinttype = {arXiv},
  pages = {1098--1105},
  issn = {1050-2947, 1094-1622},
  doi = {10.1103/PhysRevA.54.1098},
  url = {http://arxiv.org/abs/quant-ph/9512032},
  urldate = {2022-02-05},
}

@article{aaronsonQuantumComputingPostselection2005,
  title = {Quantum Computing, Postselection, and Probabilistic Polynomial-Time},
  author = {Aaronson, Scott},
  date = {2005-09-05},
  journaltitle = {Proceedings of the Royal Society A: Mathematical, Physical and Engineering Sciences},
  shortjournal = {Proc. A},
  volume = {461},
  number = {2063},
  eprint = {quant-ph/0412187},
  eprinttype = {arXiv},
  pages = {3473--3482},
  issn = {1364-5021},
  doi = {10.1098/rspa.2005.1546},
  url = {https://doi.org/10.1098/rspa.2005.1546},
  urldate = {2026-02-24}
}

@article{kokLinearOpticalQuantum2007,
  title = {Linear Optical Quantum Computing with Photonic Qubits},
  author = {Kok, Pieter and Munro, W. J. and Nemoto, Kae and Ralph, T. C. and Dowling, Jonathan P. and Milburn, G. J.},
  date = {2007-01-24},
  journaltitle = {Reviews of Modern Physics},
  shortjournal = {Rev. Mod. Phys.},
  volume = {79},
  number = {1},
  eprint = {quant-ph/0512071},
  eprinttype = {arXiv},
  pages = {135--174},
  publisher = {American Physical Society},
  doi = {10.1103/RevModPhys.79.135},
  url = {https://link.aps.org/doi/10.1103/RevModPhys.79.135},
  urldate = {2025-04-19},
}

@article{willsErrordetectedCoherenceMetrology2025,
  title = {Error-Detected Coherence Metrology of a Dual-Rail Encoded Fixed-Frequency Multimode Superconducting Qubit},
  author = {Wills, James and Haque, Mohammad Tasnimul and Vlastakis, Brian},
  date = {2025-06-18},
  eprint = {2506.15420},
  eprinttype = {arXiv},
  eprintclass = {quant-ph},
  doi = {10.48550/arXiv.2506.15420},
  url = {http://arxiv.org/abs/2506.15420},
  urldate = {2025-09-06},
  pubstate = {prepublished},
}

@article{javadi-abhariQuantumComputingQiskit2024,
  title = {Quantum Computing with {{Qiskit}}},
  author = {Javadi-Abhari, Ali and Treinish, Matthew and Krsulich, Kevin and Wood, Christopher J. and Lishman, Jake and Gacon, Julien and Martiel, Simon and Nation, Paul D. and Bishop, Lev S. and Cross, Andrew W. and Johnson, Blake R. and Gambetta, Jay M.},
  date = {2024-06-19},
  eprint = {2405.08810},
  eprinttype = {arXiv},
  eprintclass = {quant-ph},
  doi = {10.48550/arXiv.2405.08810},
  url = {http://arxiv.org/abs/2405.08810},
  urldate = {2026-02-24},
  pubstate = {prepublished},
}

@book{mackayInformationTheoryInference2003,
  title = {Information Theory, Inference and Learning Algorithms},
  author = {MacKay, David J. C.},
  date = {2003-09-25},
  eprint = {AKuMj4PN_EMC},
  eprinttype = {googlebooks},
  publisher = {Cambridge University Press},
  isbn = {978-0-521-64298-9},
  langid = {english},
  pagetotal = {694}
}

@article{reedPolynomialCodes1960,
  title = {Polynomial Codes over Certain Finite Fields},
  author = {Reed, Irving S. and Solomon, Gustave},
  date = {1960-06},
  journaltitle = {Journal of the Society for Industrial and Applied Mathematics},
  shortjournal = {Soc. Indust. Appl. Math.},
  volume = {8},
  number = {2},
  pages = {300--304},
  doi = {10.1137/0108018},
  url = {https://epubs.siam.org/doi/10.1137/0108018},
  urldate = {2026-02-26},
  langid = {english},
}

@article{grasslCodesQuantumErasure1997,
  title = {Codes for the Quantum Erasure Channel},
  author = {Grassl, Markus and Beth, Thomas and Pellizzari, Thomas},
  date = {1997-07-01},
  journaltitle = {Physical Review A},
  shortjournal = {Phys. Rev. A},
  volume = {56},
  number = {1},
  eprint = {quant-ph/9610042},
  eprinttype = {arXiv},
  pages = {33--38},
  publisher = {American Physical Society},
  doi = {10.1103/PhysRevA.56.33},
  url = {https://link.aps.org/doi/10.1103/PhysRevA.56.33},
  urldate = {2026-02-26}
}

@article{delfosseLineartimeMaximumLikelihood2020,
  title = {Linear-Time Maximum Likelihood Decoding of Surface Codes over the Quantum Erasure Channel},
  author = {Delfosse, Nicolas and Zémor, Gilles},
  date = {2020-07-09},
  journaltitle = {Physical Review Research},
  shortjournal = {Phys. Rev. Res.},
  volume = {2},
  number = {3},
  eprint = {1703.01517},
  eprinttype = {arXiv},
  eprintclass = {quant-ph},
  pages = {033042},
  publisher = {American Physical Society},
  doi = {10.1103/PhysRevResearch.2.033042},
  url = {https://link.aps.org/doi/10.1103/PhysRevResearch.2.033042},
  urldate = {2024-04-12}
}

@article{kubicaErasureQubitsOvercoming2023,
  title = {Erasure Qubits: Overcoming the {{$T_1$}} Limit in Superconducting Circuits},
  shorttitle = {Erasure Qubits},
  author = {Kubica, Aleksander and Haim, Arbel and Vaknin, Yotam and Levine, Harry and Brandão, Fernando and Retzker, Alex},
  date = {2023-11-01},
  journaltitle = {Physical Review X},
  shortjournal = {Phys. Rev. X},
  volume = {13},
  number = {4},
  eprint = {2208.05461},
  eprinttype = {arXiv},
  eprintclass = {quant-ph},
  pages = {041022},
  publisher = {American Physical Society},
  doi = {10.1103/PhysRevX.13.041022},
  url = {https://link.aps.org/doi/10.1103/PhysRevX.13.041022},
  urldate = {2026-02-26}
}

@article{derksDesigningFaulttolerantCircuits2025,
  title = {Designing Fault-Tolerant Circuits Using Detector Error Models},
  author = {Derks, Peter-Jan H. S. and Townsend-Teague, Alex and Burchards, Ansgar G. and Eisert, Jens},
  date = {2025-11-06},
  journaltitle = {Quantum},
  volume = {9},
  eprint = {2407.13826},
  eprinttype = {arXiv},
  eprintclass = {quant-ph},
  pages = {1905},
  doi = {10.22331/q-2025-11-06-1905},
  url = {https://quantum-journal.org/papers/q-2025-11-06-1905/},
  urldate = {2026-02-28},
  langid = {british}
}

@article{hanThresholdComputationCryptographic1997,
  title = {Threshold Computation and Cryptographic Security},
  author = {Han, Yenjo and Hemaspaandra, Lane A. and Thierauf, Thomas},
  date = {1997-02},
  journaltitle = {SIAM Journal on Computing},
  shortjournal = {SIAM J. Comput.},
  volume = {26},
  number = {1},
  pages = {59--78},
  publisher = {{Society for Industrial and Applied Mathematics}},
  issn = {0097-5397},
  doi = {10.1137/S0097539792240467},
  url = {https://epubs.siam.org/doi/10.1137/S0097539792240467},
  urldate = {2026-03-01}
}

@article{griffithsUnionfindQuantumDecoding2024,
  title = {Union-Find Quantum Decoding without Union-Find},
  author = {Griffiths, Sam J. and Browne, Dan E.},
  date = {2024-02-09},
  journaltitle = {Physical Review Research},
  shortjournal = {Phys. Rev. Res.},
  volume = {6},
  number = {1},
  eprint = {2306.09767},
  eprinttype = {arXiv},
  eprintclass = {quant-ph},
  pages = {013154},
  doi = {10.1103/PhysRevResearch.6.013154},
  url = {https://link.aps.org/doi/10.1103/PhysRevResearch.6.013154},
  urldate = {2024-03-07}
}

@article{gidneyStimFastStabilizer2021,
  title = {Stim: A Fast Stabilizer Circuit Simulator},
  shorttitle = {Stim},
  author = {Gidney, Craig},
  date = {2021-07-06},
  journaltitle = {Quantum},
  volume = {5},
  eprint = {2103.02202},
  eprinttype = {arXiv},
  eprintclass = {quant-ph},
  pages = {497},
  publisher = {Verein zur Förderung des Open Access Publizierens in den Quantenwissenschaften},
  doi = {10.22331/q-2021-07-06-497},
  url = {https://quantum-journal.org/papers/q-2021-07-06-497/},
  urldate = {2026-03-01},
  langid = {british}
}

@article{preskillNISQMegaquopMachine2025,
  title = {Beyond {{NISQ}}: The Megaquop Machine},
  shorttitle = {Beyond {{NISQ}}},
  author = {Preskill, John},
  date = {2025-04-29},
  journaltitle = {ACM Transactions on Quantum Computing},
  shortjournal = {ACM Transactions on Quantum Computing},
  volume = {6},
  number = {3},
  eprint = {2502.17368},
  eprinttype = {arXiv},
  eprintclass = {quant-ph},
  pages = {18:1--18:7},
  doi = {10.1145/3723153},
  url = {https://dl.acm.org/doi/10.1145/3723153},
  urldate = {2026-06-02}
}

@article{krantzQuantumEngineersGuide2019,
  title = {A Quantum Engineer's Guide to Superconducting Qubits},
  author = {Krantz, P. and Kjaergaard, M. and Yan, F. and Orlando, T. P. and Gustavsson, S. and Oliver, W. D.},
  date = {2019-06-17},
  journaltitle = {Applied Physics Reviews},
  shortjournal = {Applied Physics Reviews},
  volume = {6},
  number = {2},
  eprint = {1904.06560},
  eprinttype = {arXiv},
  eprintclass = {quant-ph},
  pages = {021318},
  issn = {1931-9401},
  doi = {10.1063/1.5089550},
  url = {https://doi.org/10.1063/1.5089550},
  urldate = {2025-04-19}
}

@article{klaverSWAPlessImplementationQuantum2026,
  title = {{{SWAP-less}} Implementation of Quantum Algorithms},
  author = {Klaver, Berend and Rombouts, Stefan M. A. and Fellner, Michael and Messinger, Anette and Ender, Kilian and Ludwig, Katharina and Lechner, Wolfgang},
  date = {2026-01-29},
  journaltitle = {Physical Review A},
  shortjournal = {Phys. Rev. A},
  volume = {113},
  number = {1},
  eprint = {2408.10907},
  eprinttype = {arXiv},
  eprintclass = {quant-ph},
  pages = {012443},
  publisher = {American Physical Society},
  doi = {10.1103/2wzk-fnhx},
  url = {https://link.aps.org/doi/10.1103/2wzk-fnhx},
  urldate = {2026-06-04}
}

@book{hughesMeasurementsTheirUncertainties2010,
  title = {Measurements and Their Uncertainties: A Practical Guide to Modern Error Analysis},
  shorttitle = {Measurements and Their Uncertainties},
  author = {Hughes, Ifan G. and Hase, Thomas P. A.},
  date = {2010-07-01},
  eprint = {zEK1DwAAQBAJ},
  eprinttype = {googlebooks},
  publisher = {Oxford University Press},
  isbn = {978-0-19-956633-4},
  pagetotal = {160}
}

\end{document}